# Role of remote interfacial phonon (RIP) scattering in heat transport across graphene/SiO$_2$ interfaces

Yee Kan Koh[1], Austin S. Lyons[2], Myung-Ho Bae,[2,3] Bin Huang,[1] Vincent E. Dorgan,[2] David G. Cahill[4], and Eric Pop,[2,5]

[1]Department of Mechanical Engineering, and Centre for Advanced 2D Materials, National University of Singapore, Singapore

[2]Department of Electrical & Computer Engineering, Micro and Nanotechnology Lab, University of Illinois, Urbana, Illinois 61801, USA

[3]Korea Research Institute of Standards and Science, Daejeon 305-340, Republic of Korea

[4]Department of Materials Science and Engineering, and Frederick Seitz Materials Research Laboratory, University of Illinois, Urbana, Illinois 61801, USA

[5]Department of Electrical Engineering and Precourt Institute for Energy, Stanford University, Stanford, CA 94305, USA

## ABSTRACT

Heat transfer across interfaces of graphene and polar dielectrics (e.g. SiO$_2$) could be mediated by direct phonon coupling, as well as electronic coupling with remote interfacial phonons (RIPs). To understand the relative contribution of each component, we develop a new pump-probe technique, called voltage-modulated thermoreflectance (VMTR), to accurately measure the change of interfacial thermal conductance under an electrostatic field. We employed VMTR on top gates of graphene field-effect transistors and find that the thermal conductance of SiO$_2$/graphene/SiO$_2$ interfaces increases by up to $\Delta G \approx 0.8$ MW m$^{-2}$ K$^{-1}$ under electrostatic fields of <0.2 V nm$^{-1}$. We propose two possible explanations for the observed $\Delta G$. First, since the applied electrostatic field induces charge carriers in graphene, our VMTR measurements could originate from heat transfer between the charge carriers in graphene and RIPs in SiO$_2$. Second, the increase in heat conduction could be caused by better conformity of graphene interfaces un-der electrostatic pressure exerted by the induced charge carriers. Regardless of the origins of the observed Δ$G$, our VMTR measurements establish an upper limit for heat transfer from unbiased graphene to $SiO_2$ substrates via RIP scattering; i.e., only <2% of the interfacial heat transport is facilitated by RIP scattering even at a carrier concentration of ~$4\times10^{12}$ $cm^{-2}$.

**TEXT**

Charge carriers in carbon nanotubes (CNTs), graphene and other 2D materials can remotely couple to surface polar phonons (SPPs) in the underlying polar substrates via a mechanism usually called remote interfacial phonon (RIP) scattering.[1-3] (We use surface polar phonons to refer to thermally excited vibrational modes near the surface that have an electric dipole moment, including e.g., surface phonon polaritons.) In RIP scattering, remote coupling between charge carriers in graphene or CNTs and SPPs in the polar substrates, spatially separated over a distance of <1 nm, is facilitated by oscillating surface electric fields created by the SPPs, and is accompanied by momentum and energy exchange.[4] RIP scattering of charge carriers has been frequently invoked to explain electrical and optical properties of graphene and CNTs, e.g., the reduction of the mobility of charge carriers in supported graphene and CNTs,[1,2] current saturation in graphene,[3,5] and nonvanishing absorption of light by graphene.[6]

In principle, with the remote coupling through RIP scattering, remote energy transfer occurs between charge carriers in graphene (or CNTs) and the dielectric substrates. While a few theoretical studies of the role of RIP scattering in heat transfer across interfaces of graphene or CNTs exist in the literature,[4, 7-10] experimental measurements[11] of heat transport due to RIP scattering are limited, due to difficulties in isolating heat transfer due to RIP scattering from heat transfer due to vibration modes (i.e., phonons). In this paper, we present a novel approach to directly measure the change of interfacial thermal conductance under electric fields, which could become a generally applicable tool to study the contribution of RIP scattering to interfacial heat transfer of many other 2D materials under varying bias conditions.

Knowledge of heat transport across graphene interfaces[12, 13] is crucial for thermal management of graphene devices. For undoped and unbiased graphene, we previously demonstrated that heat is predominantly carried across inherently decoupled graphene interfaces through transmission of acoustic phonons.[13] (Throughout this paper, we use "unbiased graphene" to refer to graphene with no electric current flowing because no lateral source-drain bias is applied.) However, under high vertical electric fields (induced by a gate, as in an operating graphene field-effect transistor), a high concentration of charge carriers is induced in the graphene channel. As a result, electrons and holes could play a significant role in heat conduction across graphene interfaces via RIP scattering.[9, 14]

In fact, this remote heat transfer via RIP scattering is thought to dominate heat dissipation from CNTs to dielectric substrates.[7, 11] Specifically, Rotkin *et al*. calculated that RIP scattering contributes thermal conductance per unit length of $g_{RIP} \approx 0.1$ W m$^{-1}$ K$^{-1}$ to interfacial heat transfer of a single-walled CNT (~1.3 nm in diameter) with a carrier concentration of 0.1 e/nm.[7] Later, Baloch *et al*. estimated from their measurements on multi-walled CNTs (~25 nm in diameter) that the contribution by RIP scattering is $g_{RIP} \approx 0.02$ W m$^{-1}$ K$^{-1}$ when driven by high source-drain biases.[11] (We discuss below our estimates related to Baloch's work, and note that in reaching their conclusion, they used a low thermal contact conductance by vibration modes of 0.004 W m$^{-1}$ K$^{-1}$, which they independently measured.[15])

Here, we experimentally demonstrate that unlike CNTs, contribution of RIP scattering to interfacial heat conduction across unbiased graphene (without current flow) and SiO$_2$ is small compared to the phonon contribution, even at high carrier concentrations. We accurately measured the change of thermal conductance of SiO$_2$/graphene/SiO$_2$ interfaces under modulated electrostatic fields and find that the thermal conductance of the graphene interfaces increases by up to ~0.8 MW m$^{-2}$ K$^{-1}$ under a vertical electrostatic field of <0.2 V nm$^{-1}$. We postulate that this (small) enhancement in heat conduction is due to either better conformity of graphene to the substrates under electrostatic pressure or an additional heat transfer channel by RIP scattering of charge carriers in graphene. In either case, we successfully establish an upper limit to heat conduction via RIP scattering; we estimate that the contribution does not exceed 1.6 MW m$^{-2}$ K$^{-1}$, which is <2% of total thermal conductance of the unbiased graphene interfaces, even at a carrier concentration of $4\times10^{12}$ cm$^{-2}$. We note that graphene in our experiments is unbiased (with no current flow), while CNTs in prior calculations[7] and experiments[11] were source-drain biased and carrying an electrical current, which could play an additional role in RIP scattering.[7, 10]

The changes of thermal conductance that we observed in this work, while still small in magnitude and thus currently not yet suitable for practical applications, could inspire a new and unconventional approach to control heat conduction in future devices, i.e., via application of electrostatic fields.

Our test structure is a dual-gated graphene field-effect transistor (GFET), see Fig. 1a and 1b. Details of sample preparation are presented in the supplementary information. We deposited graphene flakes on 90 nm SiO$_2$ on Si by micromechanical exfoliation[16] from natural graphite.

We determined the numbers of graphitic layers from the ratios of the integrated intensity of the G and 2D peaks,[17] and of the G and Si peaks[18] of the Raman spectra (Fig. 1c). We then annealed the samples in a chemical vapor deposition (CVD) furnace at 400 ºC for 35 minutes in Ar/H$_2$ mixture gas to remove the adhesive residue from the tape. After annealing, we patterned and subsequently deposited Ti/Au (1/40 nm) on the graphene as source/drain electrodes, SiO$_2$ as the gate dielectric layers and Al (90 nm) as the top gates, by e-beam lithography, e-beam evaporation, and lift-off. The thin layer of SiO$_2$ was thermally evaporated on the samples by an e-beam evaporator at ~8 × 10$^{-7}$ Torr. The thickness of the SiO$_2$ films was measured by ellipsometry; $h_{SiO2}$ = 24-25 nm.

We verified that the e-beam lithography, lift-off and evaporation processes did not damage our graphene flakes through our Raman, electrical and thermal measurements. We performed Raman spectroscopy measurements on graphene covered by evaporated SiO$_2$ and found that the Raman spectra are similar to the Raman spectra of pristine[13] graphene with reasonably small D peaks (which indicate low defect concentrations in graphene), see Fig. 1c. Figure 1d displays transistor modulation of the graphene devices using the top gates, when the bottom-gates are grounded. We obtained mobility of 1900 cm$^2$ V$^{-1}$ s$^{-1}$ and 280 cm$^2$ V$^{-1}$ s$^{-1}$ for our monolayer and bilayer graphene, respectively. We measured the thermal conductance $G$ of SiO$_2$/graphene/SiO$_2$ interfaces by the differential time-domain thermoreflectance (TDTR); details of our approach are described in Ref. 13. We find that $G \approx 41$ MW m$^{-2}$ K$^{-1}$, with an uncertainty of ~28%. The uncertainty is rather high due to low sensitivity of TDTR signals to $G$, see the sensitivity plots in Fig. S4 in the supplementary information. The measured thermal conductance is roughly half of $G$ = 83 MW m$^{-2}$ K$^{-1}$ for single SiO$_2$/graphene interface,[12] consistent with our previous conclusion that for heat conduction, graphene interfaces can be regarded as two discrete interfaces acting in series, even for monolayer graphene.[13]

We measured the change of thermal conductance of graphene interfaces under electrostatic fields by a new technique called voltage-modulated thermoreflectance (VMTR). VMTR is an extension of time-domain thermoreflectance (TDTR), a pump-probe technique to study thermal energy transport in nanostructures[19-21] and across interfaces.[13, 22, 23] In VMTR measurements, ultrashort laser pulses (≈100 fs) from a Ti:sapphire oscillator are split into a pump and a probe beams. The pump beam, modulated at a radio-frequency (rf) of $f_{pump}$ = 10 MHz, is absorbed by the metal electrodes, generating periodic heating at their surface. The probe beam is then used to monitor the periodic temperature response induced by the heating through thermoreflectance (i.e., changes of reflectance with temperature), using a photodiode and a rf lock-in amplifier. We fixed the relative time between pump and probe pulses (which is called the delay time) at $t$ = -40 ps. At this delay time, the out-of-phase signals ($V_{out}$) of the rf lock-in amplifier are inversely proportional to the thermal conductance of the samples,[24] and thus could be used to measure changes in thermal conductance induced by electrostatic fields. The rationale of choosing $t$ = -40 ps and $f$ = 10 MHz is illustrated in the sensitivity plots (Fig. S4) in the supplementary information.

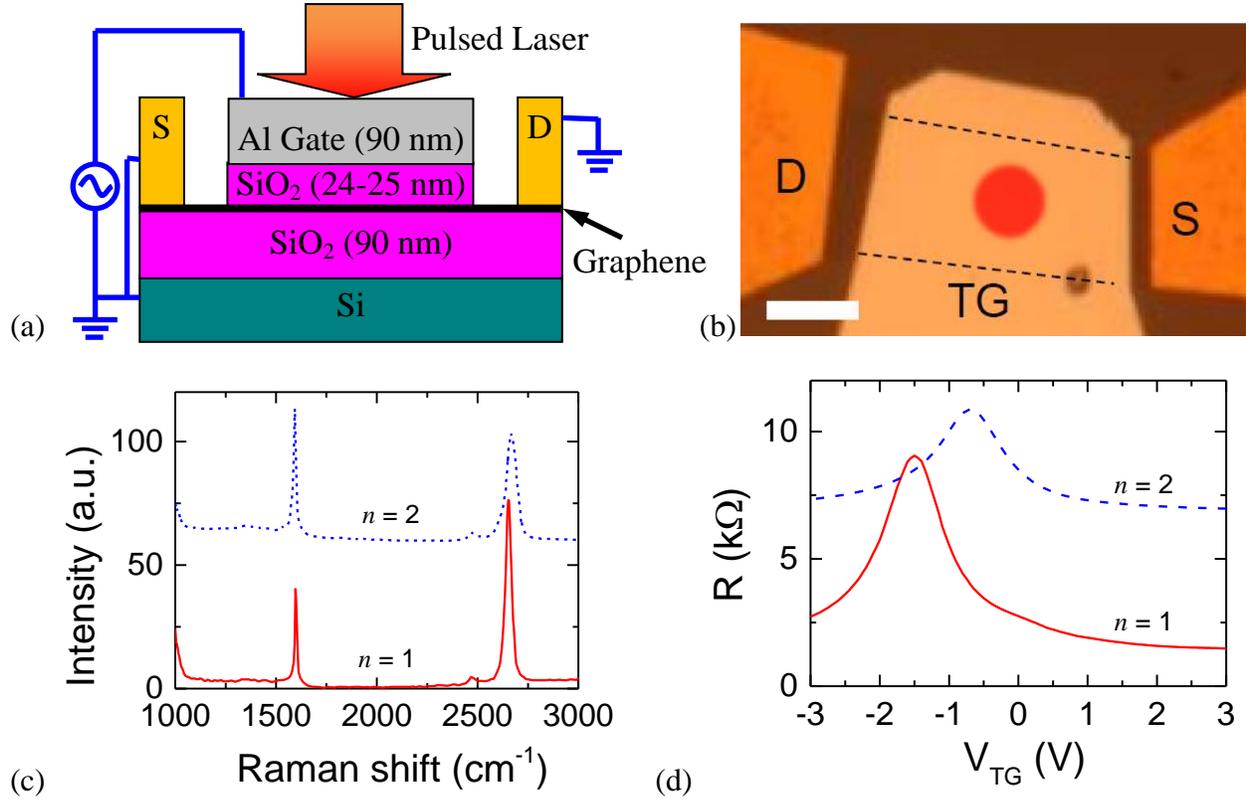

**Figure 1: (a)** Vertical cross-section of the samples (not-to-scale). An Al (90 nm) thin metal pad was patterned on a graphene flake on $SiO_2$ (90 nm) on Si substrate as the top gate of the graphene field-effect transistor (GFET). In all our measurements, the sources and drains are grounded. In voltage-modulated thermoreflectance (VMTR) measurements, the thermoreflectance signals of the top gate are monitored while the electrostatic potential of the gate is modulated. **(b)** An optical image of a monolayer GFET sample. D, S and TG indicate drain, source and top-gate electrodes. The red spot on top-gate electrode represents the laser spot for the VMTR measurement. The scale bars is 10 μm. **(c)** Raman spectra of a monolayer ($n=1$) and a bilayer ($n=2$) graphene used in our experiments. The curves are vertically shifted for the clarity. Number of graphitic layers of the graphene flakes is determined from the ratio of integrated intensity of G and 2D, and Si and G peaks as described in Refs. [17] and [18]. **(d)** Electrical resistance of our GFETs as a function of top-gate voltage ($V_{TG}$). The charge-neutral-point is located at around -1.5 V for the monolayer graphene sample and -0.75 V for the bilayer graphene sample.

To assess the effects of electrostatic fields on heat conduction, we employed 50% duty-cycle, on-off square-wave voltage modulation to the GFET top gates while keeping the Si substrates grounded, see the schematic in Fig. 1a. We set the modulation frequency at an audio-frequency (af) of $f_{TG} = 300$ Hz and changed the "on"-state gate voltage $V_{TG}$ from -5 V to 5 V. We monitored the resulting change in $V_{out}$ at frequency $f_{TG}$ ($\Delta V_{out}$) using an af lock-in amplifier. We normalized $\Delta V_{out}$ with a TDTR measurement of $V_{out}$ without gate modulation to eliminate the need to accurately measure the laser power absorbed by the metal gates. To improve the signal-to-noise of the VMTR measurements, we repeated the measurements for 40-100 times, with a time constant of 10 s for each measurement, and obtained the average values for $\Delta V_{out}$.

We note that VMTR is a substantial improvement over previous measurements (e.g., in Ref. 25) of monitoring TDTR signals as the applied voltage changes. In conventional TDTR, the accuracy in determining $\Delta V_{out}$ (or other TDTR signals) due to the applied voltage depends on the precision of TDTR measurements over time, i.e., the repeatability of TDTR measurements when the same experimental parameters are employed. The precision of TDTR measurements is usually limited by e.g., phrase drift in the rf lock-in amplifier and is typically on the order of 1%. In VMTR, however, we apply double modulations by replacing the modulation on the probe beam in TDTR with voltage modulation on the top Al transducer at an audio-frequency (e.g., 300 Hz). We then use an audio-frequency (af) lock-in amplifier to extract the TDTR responses to changes of electric fields, and thus directly and accurately measure $\Delta V_{out}$. Due to use of the af lock-in amplifier, the uncertainty of $\Delta V_{out}$ is significantly reduced from the uncertainty in the original TDTR measurements of $G$, see Fig. S5 in the Supplementary for the raw data and the uncertainty of our TDTR and VMTR measurements. The uncertainty of $\Delta V_{out}/V_{out}$ is estimated from standard deviation of the means to be $<2\times10^{-5}$, improved from the precision of conventional TDTR by more than a factor of 100. Since all other parameters (e.g., Al thickness, laser spot sizes) does not change in VMTR measurements, VMTR is not sensitive to these other parameters. Thus, VMTR is suitable to sensitively detect any small changes of heat transport under electrostatic fields.

We derived the change of the thermal conductance $\Delta G$ from our VMTR measurements; $\Delta G = G_{TG} - G_0$, where $G_{TG}$ and $G_0$ are the thermal conductance of the SiO$_2$/graphene/SiO$_2$ interfaces at the top-gate voltages of $V_{TG}$ and 0 V respectively. Since $\Delta G \ll G$, $\Delta G$ is linearly proportional to $\Delta V_{out}/V_{out}$ and can be expressed as

$$\Delta G = \alpha G_0 \frac{\Delta V_{out}}{V_{out}} \mathrm{sgn}(V_{TG}) \qquad (1)$$

where $\alpha = (V_{out}/G_0)(\partial G_0/\partial V_{out})$ is a dimensionless proportionality constant derived from the thermal model[24] for TDTR measurements; $\alpha \approx 10$ for our GFET samples. $\mathrm{sgn}(V_{TG})$ in Eq. (1) is the sign of the applied "on"-state gate voltage and it accounts for the flip of sign of $\Delta V_{out}$ across $V_{TG} = 0$ V. At negative $V_{TG}$, the af lock-in amplifier in our setup reads a negative value of $\Delta V_{out}$ for an increase of $V_{out}$, due to 180° out-of-phase of the gate voltage and the corresponding $V_{out}$. Thus, a negative value of $\Delta V_{out}$ at a negative $V_{TG}$ denotes an increase of $V_{out}$.

In Fig. 2a, we plot our VMTR measurements on the GFETs as a function of the electrostatic field across the SiO$_2$/graphene/SiO$_2$ interface during the "on"-state of gate modulation, $E_{TG}$. We estimated $E_{TG}$ from the "on"-state gate voltage $V_{TG}$ using $E_{TG} = V_{TG}/h_{SiO2}$. Our $\Delta V_{out}/V_{out}$ measurements at large $E_{TG}$ are two orders of magnitude larger than the experimental uncertainty, and thus are not an experimental artifact. We find that the measured $\Delta V_{out}/V_{out}$ are mostly positive at positive $E_{TG}$ and negative at negative $E_{TG}$, suggesting that the thermal conductance increases with gate voltages of both signs. Also, we observe that the VMTR measurements differ significantly between samples, with no clear systematic dependence on the number of graphitic layers, see Fig. 2a. For example, in Fig. 2a, the VMTR signals are very different for 2 bilayer graphene samples, and thus indicating that there must be some other important factors (such as the degree of conformity, roughness etc.) which plays a major role. We are unsure of what factors contribute to the observed difference in the VMTR measurements.

We postulate that the increase of interfacial thermal conductance could be due to either additional heat transfer channel by induced charge carriers in graphene or electrostatic pressure exerted by the induced charge carriers. To test the postulates, we fit the VMTR measurements in Fig. 2a using the corresponding dependence on the carrier concentrations in graphene. If the VMTR measurements are due to additional heat transfer channel by the induced charge carriers in graphene, e.g., via remote interfacial phonon (RIP) scattering[1, 7, 11, 14] of charge carriers in graphene by surface polar phonons (SPPs) in $SiO_2$, the change of the thermal conductance of the $SiO_2$/graphene/$SiO_2$ interface ($\Delta G$) is approximately proportional to the change of sum of the concentrations of electrons ($n$) and holes ($p$) in graphene, as long as the charge concentrations are sufficiently low and the SPPs are essentially unscreened. Mathematically, $\Delta G \sim \Delta(n+p)$, where $\Delta(n+p) = (n+p)_{TG} - (n+p)_0$ and subscripts "TG" and "0" denote values under top gate voltages of $V_{TG}$ ("on"-state) and 0 V ("off"-state) respectively. We estimate[26] that $n+p = \sqrt{n_{cv}^2 + 4n_0^2}$. Here, $n_{cv} = |n-p| = (C_{ox}/q)(|V_G-V_0|)$ is the carrier concentrations induced by the applied gate voltage $V_G$, $V_0$ is the Dirac voltage, $C_{ox}$ is the top-gate oxide capacitance, $q = 1.6 \times 10^{-19}$ C is the elementary charge, and $n_0$ is the carrier density due to thermal generation and electrostatic spatial inhomogeneity in $SiO_2$. We derived $V_0$ from the transistor modulation measurements for two of our GFETs, see Fig. 1d, and obtained $V_0$ for the remaining two GFETs by fitting our VMTR measurements. We assume a dielectric constant of 3.9 for the top-gate oxides, see the justification in the supplementary information, and obtain $C_{ox} \approx 1.4 \times 10^{-3}$ F m$^{-2}$ for our GFETs. We estimate $n_0 = $ 3.3-6.8×10$^{11}$ cm$^{-2}$ using the measured or fitted values of the Dirac voltage, following the procedures outlined in Ref. 26.

Using the estimated $\Delta(n+p)$, we fitted our VMTR measurements assuming $\Delta G \sim \Delta(n+p)$, see the dashed lines in Fig. 2a. The fits are reasonably good except for the largest $\Delta V_{out}/V_{out}$ at $E_{TG}$ = -0.18 V nm$^{-1}$. We then derived $\Delta G$ from our VMTR measurements using Eq. (1) and plot $\Delta G$ as a function of $\Delta(n+p)$ in Fig. 2b. We note that the derived $\Delta G$ has much higher uncertainty than our VMTR measurements in Fig. 2a, due to the high uncertainty of $G_0$ in Eq. (1). The measured $\Delta G$ is independent of types of charge carriers (i.e., electrons or holes) induced in graphene, which is consistent with the ambipolar nature of graphene. The largest $\Delta G$ = 0.8 MW m$^{-2}$ K$^{-1}$ occurs at $E_{TG}$ = -0.18 V nm$^{-1}$, or equivalent $\Delta(n+p)$ = 3.7×10$^{12}$ cm$^{-2}$. We thus can place an upper bound on the remote energy transfer between charge carriers in unbiased graphene and SPP modes in SiO$_2$ via RIP scattering; $G_{RIP}$ < 1.6 MW m$^{-2}$ K$^{-1}$ for a single graphene/SiO$_2$ interface at $n+p \approx 4\times 10^{12}$ cm$^{-2}$. (We multiply the measured $\Delta G$ by a factor of two because there are two graphene/SiO$_2$ interfaces in our sample.)

The upper bound value of $G_{RIP}$ derived from our VMTR measurements is an order of magnitude lower than heat conduction by phonons across graphene/SiO$_2$ interfaces.[12, 13, 27] (However, we note that RIP heat conduction has been predicted to be up to an order of magnitude higher with Al$_2$O$_3$ substrates.[4]) In addition, our findings are also in contrast to the apparently dominant contribution of RIP scattering to the interfacial heat conduction of CNTs on SiO$_2$ substrates, predicted by calculations of Rotkin *et al*.[7] for single-wall CNTs and the experiments of Baloch *et al*. for multi-wall CNTs.[11] Rotkin *et al*. calculated that, for single-walled CNTs on SiO$_2$, the thermal conductance per unit length due to RIP scattering is $g_{RIP} \approx 0.1$ W m$^{-1}$ K$^{-1}$ when the doping level is 0.1 e/nm (i.e., ~2.5 × 10$^{12}$ cm$^{-2}$) and the lateral field is >2 V μm$^{-1}$. Baloch *et al*. estimated from their measurements on metallic multi-walled CNTs that remote heat transfer via RIP scattering is ~5 times larger (i.e., 84 % vs. 16 %) than heat dissipation via vibration modes that they previously measured ($g \approx 0.004$ W m$^{-1}$ K$^{-1}$).[15] We thus approximate $g_{RIP} \approx 0.02$ W m$^{-1}$ K$^{-1}$ for heat transfer by RIP scattering in the experiments of Baloch et al.[11]

To compare these CNT results with our VMTR measurements on graphene interfaces, we assume a width of overlap between the CNTs and the substrates of ~1 nm for the CNTs in the calculations by Rotkin et al. and of ~5 nm for the CNTs in the experiments by Baloch et al. Using these estimated footprints, the average thermal conductance per unit area by RIP scattering across the CNT/dielectric interfaces is derived as 100 MW m$^{-2}$ K$^{-1}$ (Rotkin et al.) and 4 MW m$^{-2}$ K$^{-1}$ (Baloch et al.), respectively. It is interesting to note that the upper limit of the RIP contribution derived from our measurements (1.6 MW m$^{-2}$ K$^{-1}$) is on the same order of magnitude as that derived from measurements by Baloch et al. The difference is actually in heat conduction by phonons across interfaces of graphene and CNTs, ~50 MW m$^{-2}$ K$^{-1}$ for graphene interfaces[12, 13, 27] and ~0.004 W m$^{-1}$ K$^{-1}$ for the multi-wall CNT interfaces.[15]

There are two possible explanations for the different behaviour in graphene and CNTs. First, unlike graphene that mostly conforms to the substrates,[28] CNTs do not conform to the substrates and thus the actual contact area for CNTs and the substrates is minute, especially if the substrate roughness is comparable to the nanotube diameter. The minute contact area diminishes heat conduction by phonons (i.e., vibration modes). Since RIP scattering across nanometer-sized gaps and voids is still substantial,[1] the effective contact area (i.e., footprint) for the remote energy transfer could be significantly larger for non-conformal CNTs. The larger effective contact area enhances the relative contribution of remote energy transfer due to RIP scattering in CNTs. Another key difference is that our measurements are for unbiased graphene, without current flowing, whereas the CNT predictions and measurements assumed significant current flow.[7,11] RIP scattering could be an order of magnitude stronger when the charge carriers are driven under high electric field,[7] due to e.g., excitation of high-energy SPP modes in the polar substrates by hot electrons in graphene or CNTs. Thus, the low-energy of charge carriers in our unbiased graphene could contribute to low $G_{RIP}$ compared to those measured in biased CNTs.[11] Ref. 10 has predicted that for even for graphene, a lateral field >0.5 V/μm could lead to substantial scattering between graphene and RIPs (on $SiO_2$), however this is a regime that must be examined in future experiments.

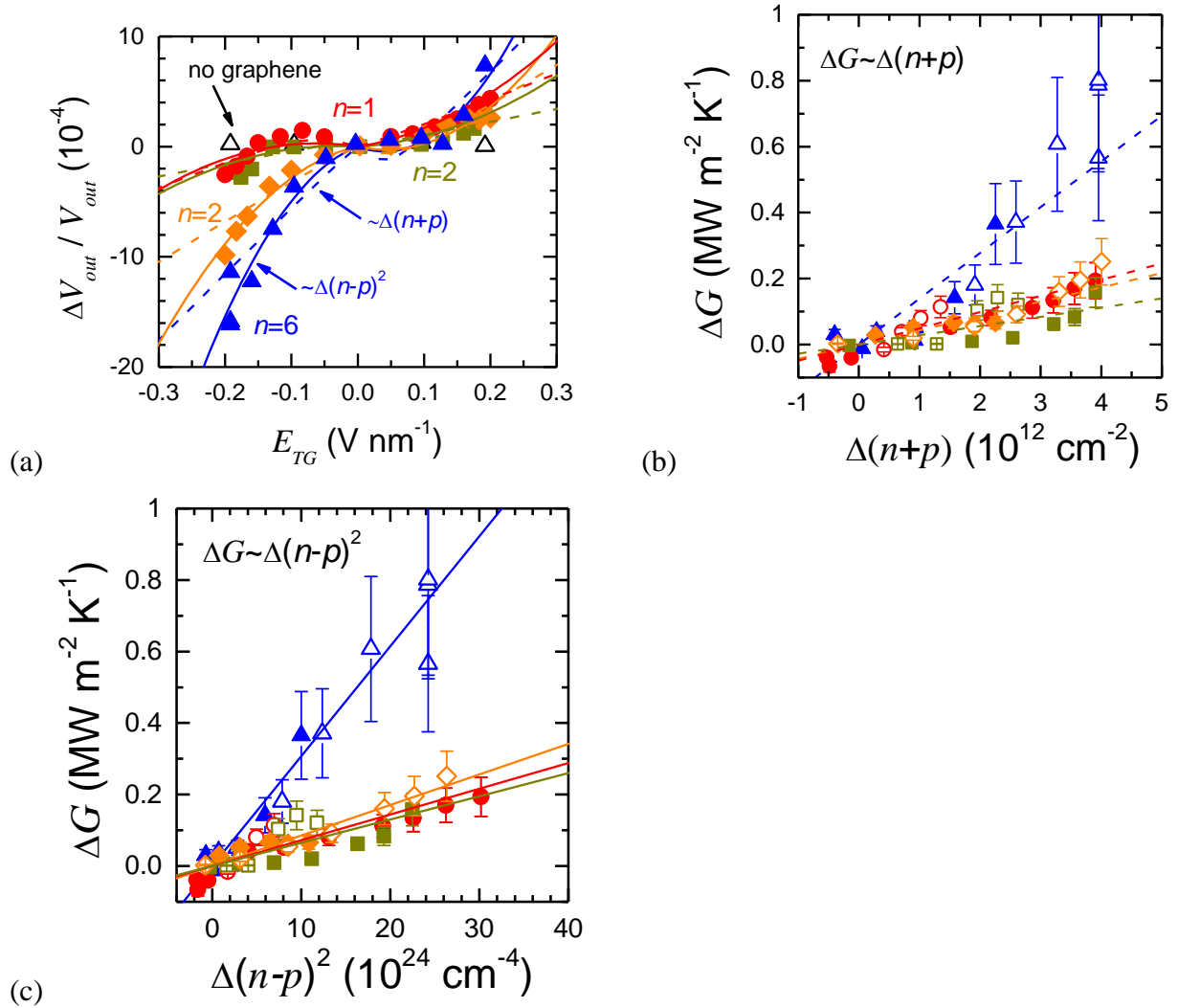

**Figure 2: (a)** Voltage-modulated thermoreflectance (VMTR) measurements ($\Delta V_{out}$) normalized by the out-of-phase signal ($V_{out}$) without top gate modulation, for a monolayer graphene (circles), two bilayer graphene (squares and diamonds) and a few-layer graphene (triangles) GFETs, as a function of the "on"-state electrostatic field ($\Delta E_{TG}$) applied to the Al top gate of the GFETs. Open triangles are VMTR measurements on the few-layer graphene sample on a region without the graphene flake. We estimate the uncertainty from standard deviations of the mean of the measurements; they are smaller than the size of the symbols. The VMTR measurements are fitted assuming that $\Delta G \sim \Delta(n+p)$ (dashed lines) and $\Delta G \sim \Delta(n-p)^2$ (solid lines), where $n$ and $p$ are electron and hole concentrations, $\Delta G = G_{TG} - G_0$, $\Delta(n+p) = (n+p)_{TG} - (n+p)_0$ and $\Delta(n-p)^2 = (n-p)^2_{TG} - (n-p)^2_{TG}$, the subscript "$TG$" and "0" represent properties at gate voltages of $V_{TG}$ and 0 V. **(b) and (c)** $\Delta G$ of SiO$_2$/graphene/SiO$_2$ interfaces, as a function of **(b)** $\Delta(n+p)$ and **(c)** $\Delta(n-p)^2$. Solid symbols are for the case that electrons are the dominant carrier ($n>p$), and open symbols for holes as the dominant carriers ($n<p$).

We further compare our VMTR measurements to theoretical calculations of interfacial thermal conductance via RIP scattering ($G_{RIP}$) in Fig. 3a. To have a fair comparison, we approximate $G_{RIP}$ during the "off"-state of our VMTR measurements, $G_{RIP,0}$, from the charge concentrations $(n+p)_0$ using the fits in Fig. 2b. We then derive $G_{RIP} = \Delta G + G_{RIP,0}$ from our VMTR measurements in Fig. 2b, and plot $G_{RIP}$ of the monolayer and the few-layer graphene samples as a function of $(n+p)_{TG}$ in Fig. 3a. In the same figure, we plot calculations of two models by Ref. 4 as solid lines. In the first model (labeled "SPP"), the authors assume that the SPPs in the oxide are electrostatically screened but are essentially decoupled from the charge carriers in graphene. With this assumption, the thermal conductance increases monotonically with carrier concentrations, but the dependence is rather weak due to enhanced electrostatic screening at high carrier concentrations, see Fig. 3a. In the second model (labeled "IPP"), the authors assume that the SPPs in the oxide are strongly coupled to the plasmons in graphene forming interfacial plasmon-phonons (IPPs). The authors predicted a stronger thermal conductance at low carrier concentrations because the plasmon motion is out of phase with the SPP modes and thus do not screen the SPP modes.[4] We find that predictions of both models contradict the stronger than linear dependence on carrier concentration that we observe in our VMTR measurements, see Figs. 2a, 2b and 3a.

The second possible explanation of the observed increase in $\Delta G$ at high gate voltages is that heat flow is enhanced by the electrostatic force exerted by induced charge carriers in graphene, resembling tuning of phononic heat conduction by hydrostatic pressure.[29, 30] If this is the case, we expect $\Delta G \sim \Delta(n-p)^2$, where $\Delta(n-p)^2 = (n-p)^2_{TG} - (n-p)^2_0$. $(n-p)$ are derived from the gate voltages as previously described. Using this dependence, we fit our VMTR measurements, see the solid lines in Fig. 2a. We also plot the derived $\Delta G$ from our VMTR measurements as a function of $\Delta(n-p)^2$ in Fig. 2c. We find that our VMTR measurements are slightly better fitted using the assumption that $\Delta G \sim \Delta(n-p)^2$ than the assumption that $\Delta G \sim \Delta(n+p)$, especially at high gate voltages as shown in Figs. 2a and 2c.

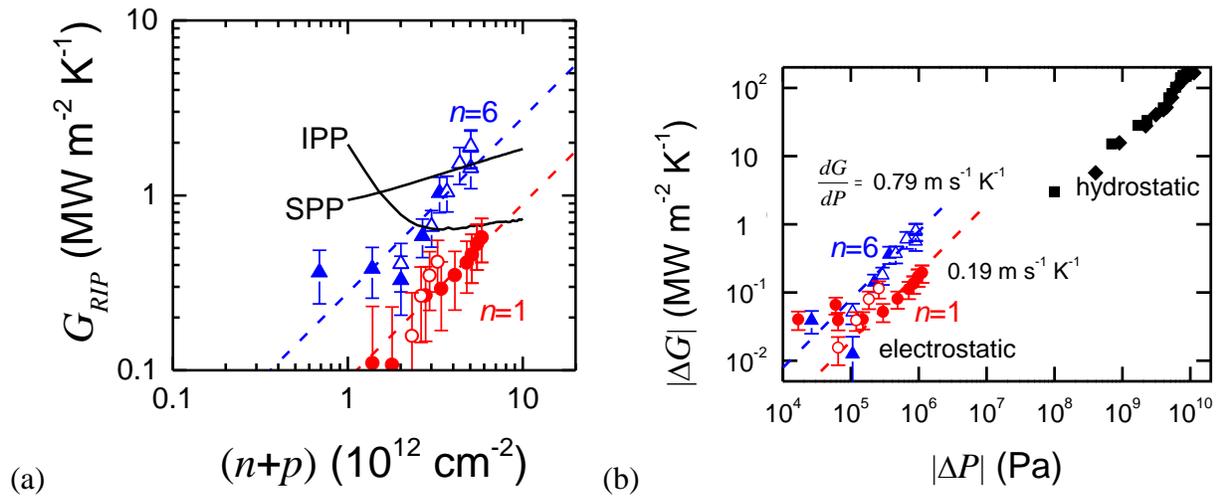

**Figure 3: (a)** The thermal conductance ($G_{RIP}$) across graphene/SiO$_2$ interfaces due to remote interfacial phonons (RIP) scattering, estimated from our VMTR measurements of a monolayer graphene (circles, this work) and a few-layer graphene (triangles, this work), with the assumption that the VMTR measurements are due to RIP scattering. The estimated $G_{RIP}$ is compared with calculations (solid lines) by Ref. [4] taking into account either static screening of surface polar phonons in SiO$_2$ (labeled "SPP") or hybridization of plasmon in graphene and SPPs in SiO$_2$ (labeled "IPP"). The dashed lines are fitted lines in Fig. 2. **(b)** Increase in the thermal conductance ($\Delta G$) of SiO$_2$/graphene/SiO$_2$ interfaces of a monolayer graphene (circles, this work) and a few-layer graphene (triangles, this work), as a function of increase in the electrostatic pressure, as-suming that the VMTR measurements are due to better conformity of the graphene under electrostatic pressure. The measured $\Delta G$ is compared to increase in the thermal conductance of Al/graphene/SiC (squares) and Al/graphene/SiO$_2$ interfaces (diamonds) due to hydrostatic pressure [29]. The dashed lines are dG/dP of 0.86 and 0.16 m s$^{-1}$ K$^{-1}$ respectively.

In Fig. 3b, we compare the measured $\Delta G$ of SiO$_2$/graphene/SiO$_2$ interfaces due to change of electrostatic pressure to the change of thermal conductance of Al/graphene/SiC and Al/graphene/SiO$_2$ interfaces due to change of hydrostatic pressure.[29] We estimate the change of electrostatic pressure ($\Delta P$) induced by the gate voltages from $\Delta P = [q^2\Delta(n–p)^2]/(2\varepsilon)$, where $\varepsilon = 3.5\times10^{-11}$ F m$^{-1}$ is the permittivity of SiO$_2$ and $q$ is the elementary charge constant. We plot $\Delta G$ as a function of $\Delta P$ and find a linear dependence of $\Delta G$ on $\Delta P$, with a slope of $dG/dP$ = 0.2-0.8 m s$^{-1}$ K$^{-1}$. The dependence on $\Delta P$ is stronger than dependence of graphene interfaces under high hydrostatic pressure[29] by a factor of ~10, see Fig. 3b.

We consider a few possible mechanisms that could engender increase in heat conduction across graphene interfaces under electrostatic pressure. First, the observed $\Delta G$ could be caused by strengthening of interface stiffness of graphene/SiO$_2$ interfaces under electrostatic pressure, similar to enhanced interface stiffness under high hydrostatic pressure.[29] With higher interface stiffness, more high-frequency phonons could transmit through the weak graphene interfaces, and thus result in higher thermal conductance. However, we find this explanation unsatisfactory due to significantly stronger dependence on $\Delta P$ observed in our measurements, compared to that under high hydrostatic pressure. The huge difference is unlikely if the underlying mechanism were the same. Second, the observed $\Delta G$ could be due to enhanced near-field radiative heat transfer across smaller gaps between graphene and SiO$_2$ under electrostatic pressure.[31] This explanation is nonetheless rather unlikely as well, since calculations[31] indicate that near-field radia-tive thermal conductance across a 1 nm gap is only ~0.05 MW m$^{-2}$ K$^{-1}$ at room temperature, two orders of magnitude smaller than our measurements.

Lastly, since graphene is only partially conformal to the SiO$_2$ surfaces,[27, 28] graphene and SiO$_2$ might elastically deform under electrostatic pressure, resulting in larger contact areas between graphene and SiO$_2$ and thus higher thermal conductance.[27, 32, 33] Under elastic deformation of graphene, prior analysis[33] suggests that when $\Delta P$ is small, $dG/dP = G_0/(3P_0)$, where $G_0$ and $P_0$ are the thermal conductance and pressure without the applied electrostatic fields, respectively. For our VMTR measurements, $G_0 \approx 41$ MW m$^{-2}$ K$^{-1}$ and $P_0 = 10^7$ Pa; thus $dG/dP \approx 1.4$ m s$^{-1}$ K$^{-1}$, on the same order of magnitude as our measurements. Moreover, the difference in $dG/dP$ that we observe in the VMTR measurements of our samples could be readily explained by different degrees of conformity of the graphene flakes. We thus propose the enhanced conformity of graphene to SiO$_2$ substrates under electrostatic pressure as the possible explanation to our measured $\Delta G$.

In summary, we observed an enhancement of heat conduction across graphene/SiO$_2$ interfaces under electrostatic fields through our accurate VMTR measurements, and thus experimentally establish an upper limit for heat conduction by RIP scattering across interfaces of graphene and SiO$_2$. Unlike predicted for biased CNTs, we find that the contribution of RIP scattering is rather insignificant (<2%) for graphene/SiO$_2$ interfaces, with unbiased graphene. We propose two possible explanations to the measured (but small) increase in the thermal conductance of graphene interfaces, i.e., additional heat transfer channel via RIP scattering and better conformity of graphene to the SiO$_2$ substrates under electrostatic pressured exerted by charge carriers in graphene. We argue that this enhancement of heat conduction under electrostatic fields should be generic to 2D materials that do not fully conform to the polar substrates, and could be signifi-cantly larger for graphene on rough substrates with poor conformity,[27] for substrates like $Al_2O_3$,[4] or when the 2D materials carry a large current driven by high source-drain biases, where the RIP heat transfer has been predicted to be enhanced.[4, 7, 10] Thus, our results could lead to a convenient approach to control phonon transport in future phononic thermal devices.

## ACKNOWLEDGEMENTS

This work was supported by SMF-NUS Research Horizons Award, NUS Young Investigator Award (Y.K.K.), DOE grant DE-FG02-07ER46459, ONR MURI grant N00014-07-1-0723 (D.G.C.), AFOSR grant FA9550-14-1-0251, NSF EFRI 2-DARE grant 1542883, and the NSF award 13-46858 (A.S.L., V.E.D., M.H.B. and E.P.), and National Research Foundation of Korea (NRF-2015R1A2A1A10056103, 2016R1A5A1008184) (M.H.B.). Sample characterization was carried out in part in the Frederick Seitz Materials Research Laboratory Central Facilities, University of Illinois, which are partially supported by the U.S. Dept. of Energy under grants DE-FG02-07ER46453 and DE-FG02-07ER46471.

## SUPPORTING INFORMATION

Fabrication process; sample images and Raman spectra; electrical characterization; sensitivity plots; comparison of raw data of TDTR and VMTR